\documentclass[twocolumn,twoside,slac_two]{revtex4}
\usepackage{graphicx}
\usepackage{fancyhdr}

\newcommand{\be}{\begin{equation}}
\newcommand{\ee}{\end{equation}}
\newcommand{\ba}{\begin{array}}
\newcommand{\ea}{\end{array}}
\newcommand{\order}{{\cal O}}

\begin{document}

\title{Lattice QCD and Flavour Physics}


\author{E. G\'amiz}
\affiliation{University of Illinois, Urbana, USA}

\begin{abstract}

I report on recent progress and future prospects for lattice QCD 
calculations relevant for flavour physics and CP violation. I will focus on 
lattice studies that incorporate realistic vacuum polarization effects, 
i.e., with $n_f=2+1$ sea quarks.
\end{abstract}

\maketitle

\thispagestyle{fancy}

% References should be done using the \cite \ref, and \label commands

\section{Introduction}

Theoretical calculations of nonperturbative QCD effects are essential 
to make full use of the results generated by the 
experimental flavour program. The combination of theory and experiment 
can be used to extract the values of the elements of the 
Cabibbo-Kobayashi-Maskawa (CKM) matrix, look for new physics (NP) 
and constrain beyond the Standard Model (BSM) theories. 
In order to be useful for achieving these goals, the theoretical 
calculations must have errors of 
the same order of the experimental errors. This requires uncertainties 
at the few percent level.

Lattice QCD is a nonperturbative formulation of QCD based 
only on first principles. It also provides a quantitative calculation 
methodology, which has become a precise tool capable of providing some of the 
accurate determinations needed by phenomenology. The processes that 
can be analyzed with current lattice QCD methods in 
a precise way are those with stable 
(or almost stable) hadrons and no more than one hadron in the initial (final) 
state. Those processes include leptonic and semileptonic decays and 
neutral meson mixing. 

Accuracy in lattice calculations requires control over all the sources 
of systematic error. In particular, it is essential to take into account 
vacuum polarization effects in a realistic way, i.e., including up, down 
and strange sea quarks on the gauge configurations' generation. The up and 
down quarks are usually taken to be degenerated, so those simulations 
are referred to as $n_f=2+1$. The vacuum polarization effects 
were almost always neglected in old lattice calculations due to limited 
computational power. This is known as the quenched approximation and  
introduces an uncontrolled and irreducible error, which can be as 
large as 10-30\% \cite{ratiopaper}. Simulations with $n_f=2$ sea quarks 
are still missing part of the vacuum polarization effects and the 
associated systematic error is hard to estimate without repeating 
the calculation with $n_f=2+1$ sea quarks. 

Another important source of systematic error is associated with the fact that 
current simulations are unable to simulate up and down quarks as light 
as the physical ones. The way of connecting lattice results to the physical 
world in a model independent way is by extrapolating those results 
using the guide of Chiral Perturbation Theory (ChPT). In order 
for this extrapolation to have controlled 
errors and be realistic, simulations must be performed for a range 
of light sea quark masses smaller than $m_s/2$. 

Other systematic errors that must be included in any realistic 
lattice analysis are discretization, finite volume and 
renormalization effects. Discretization effects can be estimated 
by power counting, but this estimate must be explicitly tested 
by performing the calculation at several values of the lattice spacing. 
Finite volume effects can be estimated by repeating the calculation 
at several volumes and/or using ChPT techniques.

\begin{figure}[h]
\hspace*{-1.2cm}\label{ratioplot}
\begin{center}
$N_f=0$ \hspace*{3.3cm}$N_f=2+1$ 
\includegraphics[width=0.4\textwidth,angle=-90]{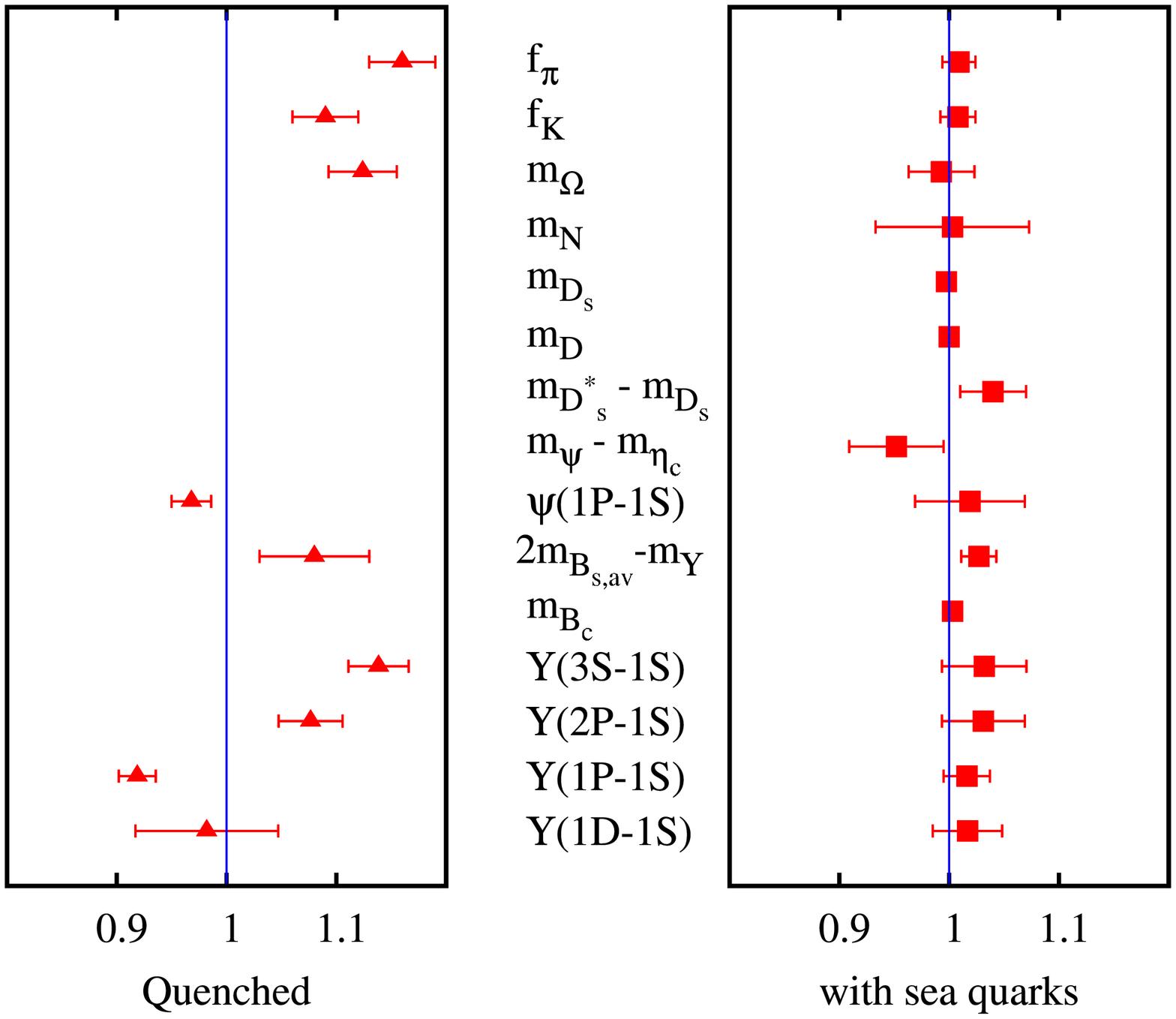}
\caption{Lattice QCD results divided by experimental results in the 
quenched approximation (left panel) and with $n_f=2+1$ sea quarks 
(right panel). Lattice values are obtained using improved staggered 
actions to describe sea quarks and $u$, $d$, $s$ and $c$ valence quarks, 
and NRQCD to simulate the $b$ valence quarks. The plot is an updated 
version of the one in \cite{Christinelp}.}
\end{center}
\end{figure}

It is important to establish the validity of lattice methods by 
comparing its predictions against 
well known experimental quantities. Figure \ref{ratioplot} shows the 
ratio between lattice results and experimental measurements for 
different decay constants, hadron masses and mass splittings. 
The agreement between lattice QCD and experiment is remarkable once 
vacuum polarization effects are included in a realistic way. This  
gives us confidence in lattice techniques. 

In the next Sections I will discuss processes relevant for the experimental 
flavour physics program for which lattice QCD can provide accurate 
determinations of the nonperturbative inputs. I will thus restrict 
my discussion to calculations with all sources of systematic error addressed. 
Among other things, that means that I will focus on 
simulations with $n_f=2+1$ sea quarks.

\subsection{Lattice fermion formulations}

\label{formulations}

The inclusion of quarks in the lattice QCD action, besides being expensive 
in computing time, has associated several difficulties. One of them is 
the so-called doubling problem, that consists of the fact that when one 
discretizes the QCD quark action in the most straightforward way, the 
naive discretization, there appear 15 additional unphysical tastes 
for any continuum flavour. Several methods exist to deal with the 
doubling problem. The most popular choice is (improved versions of) 
the Wilson action \cite{Wilson}, 
that solves the doubling problem by adding a dimension five operator with 
the cost of breaking chiral symmetry. Other fermion formulations 
keep an exact, Overlap \cite{overlap}, or almost exact, Domain Wall \cite{DW}, 
chiral symmetry paying the price of significantly complicating the 
operator structure and thus increasing the computing cost. The recent 
progress on the development of algorithms however, is starting to make 
feasible to perform unquenched calculations with those formulations.

Staggered fermions have good chiral properties and are computationally 
more efficient than any other light fermion formulation. The downside 
is that the doubling problem is not completely eliminated but reduced 
to four tastes for any continuum flavour. In unquenched simulations 
the extra tastes are eliminated by taking the root of the fermion 
determinant in the generation of configurations. This procedure is 
the focus of intense scrutiny by the lattice community and there has been 
considerable progress in our understanding of the issues involved 
during the past years \cite{fourthroot}. To date 
there is no proof of the correctness or incorrectness of the method, 
but all the tests performed on dynamical simulations taking the root 
of the determinant have given evidence of controlled effects which 
disappear in the continuum limit.

The Asqtad \cite{asqtad} is the staggered formulation most  
widely used. It is improved with respect to the original  
staggered action so leading discretization effects are removed 
at tree-level. Remaining errors are therefore $\order(\alpha_s a^2)$ 
and $\order(a^4)$. That is the fermion action that the MILC 
collaboration is using in the generation of the $n_f=2+1$ configurations, 
which are employed  
in several of the lattice calculations referred to in the next sections. 
An even more improved staggered action is the recently developed 
HISQ (highly improved staggered quark) action \cite{hisq}. It reduces 
the remaining $\order(\alpha_s a^2)$ errors coming from 
taste-changing effects in the Asqtad action by roughly a factor of 
three.

For heavy quarks, charm and bottom, discretization errors coming in powers of 
the mass in lattice units, $am_Q$, are not negligible at typical  
lattice spacings. 

The HISQ action however incorporate corrections 
that remove the dominant $a m_Q$ effects so the leading mass 
discretization effects are $\order(\alpha_s(am_c)^2)$ and $\order(am_c)^4$. 
This action can thus be used in a very effective way to describe charm 
quarks if the lattice spacing is small enough\footnote{The MILC ensembles 
used by the HPQCD collaboration in its HISQ studies have small 
enough lattice spacings so the heavy discretization errors for the charm quark 
are under control.}. 

To describe bottom quarks, however, an effective field theory framework is 
more adequate. Some implementations of the effective field methods 
are lattice heavy quark effective theory (HQET) (whose leading term 
is the static approximation), non-relativistic QCD (NRQCD) and the 
Fermilab approach. The lattice NRQCD Lagrangian is obtained by discretizing 
the non-relativistic expansion of the continuum Dirac Lagrangian.  
The particular lattice NRQCD action used in the HPQCD calculations  
of $B$ decay constants and $B^0$ mixing 
parameters described in next sections is improved through 
${\order(1/M^2)}, \,{\order(a^2)}$ and leading 
relativistic ${\order(1/M^3)}$ \cite{NRQCD}.  The action parameters 
are fixed via heavy-heavy simulations, in particular the valence 
$b$ quark mass is tuned to give the physical value of the $\Upsilon$ mass. 

The Fermilab action \cite{kkm} is obtained following a different approach. It  
starts with an improved relativistic Wilson action \cite{Wilson}, 
which has the same heavy quark limit as QCD. With the Fermilab 
interpretation in terms of HQET\cite{kkm}, this action can be used to describe 
heavy quarks without errors that grow as $(am_Q)^n$. One of the advantages 
of this approach is that it can be used for both charm and bottom quarks. 
The errors associated with the use of the Fermilab action mentioned in this 
paper, used by the FNAL/MILC collaboration, are 
$\order\left(\alpha_s\Lambda/m_Q\right)$ and 
$\order\left(\Lambda/m_Q\right)^2$.

\section{Leptonic decays}

\label{leptonic}

The lattice determination of pseudoscalar decay constants, 
together with experimental measurements of pseudoscalar 
leptonic decay widths, can be used to extract the value 
of the CKM matrix elements involved in the process  
\be
\Gamma(P_{ab}\to l \nu) = ({\rm known\,factors})\, f_P^2\vert V_{ab} \vert^2
\, .
\ee 
On the other hand, for decay constants well determined experimentally, 
those for which the CKM matrix elements involved are known with a good 
precision and experimental measurements are accurate, the comparison with 
lattice calculations can be used as a test of the theory used to make 
the theoretical prediction. 

The calculation of decay constants on the lattice is done using 
a simple matrix element,
\be
\langle 0 \vert \bar q \gamma_{\mu} \gamma_5 
c\vert P(p)\rangle = i f_{P}p_{\mu}\,, 
\ee
so they can be obtained with a very good accuracy. In the next 
subsections I will summarize the status of unquenched lattice 
calculations of charm and bottom decay constants. 
The current status of the lattice calculations of $f_K/f_\pi$, 
from which one can extract the value of the CKM matrix element 
$\vert V_{us} \vert$, was reviewed by F.~Mescia in this conference 
\cite{mesciaFPCP08}.

\subsection{$D^+$ and $D_s$ decay constants}

For the decays constants in the charm sector $f_{D^+}$ and $f_{D_s}$ 
there are lattice results available from two groups 
with $n_f=2+1$ sea quarks, the FNAL/MILC \cite{fDfnal} and the 
HPQCD \cite{fDhpqcd} collaborations, to compare against 
experiment. Both use configurations 
generated by the MILC collaboration for three different values of 
the lattice spacing, $a=0.15~fm$, $a=0.12~fm$ and $a=0.9~fm$. 
The main difference between the two collaborations is the treatment 
of the valence quarks. While HPQCD uses the HISQ 
action for all the valence quarks, FNAL/MILC uses staggered Asqtad for 
the light quarks (up, down and strange) and the Fermilab action for 
the charm quark. The HPQCD collaboration has partially conserved currents, so 
they can extract the value of the decay constant without any 
renormalization. The FNAL/MILC collaboration needs to renormalize its 
currents, but they do it in a partially non-perturbative way that 
generates very small errors, around $1.5\%$.

The values for $f_{D^+}$, $f_{D_s}$ and the ratio of both 
quantities, together with the new CLEO-c experimental 
results presented in this conference \cite{stoneFPCP08}, are collected 
in Table \ref{tablefD}. 
\begin{table}[h]
\begin{center}
\begin{tabular}{|c c c c c|}
\hline
\hline
Group & Reference & $f_{D^+}$ & $f_{D_s}$ & $f_{D_s}/f_{D^+}$ \\
\hline 
experiment & \cite{stoneFPCP08} & \begin{tabular}{c}(\emph{CLEO-c})\\
205.8(8.9)\end{tabular}
  & \begin{tabular}{c}(\emph{average})\\ 269.6(8.3) \end{tabular} & 
1.31(7)\\ 
HPQCD & \cite{fDhpqcd} &  207(4) & 241(3) & 1.164(11) \\
FNAL/MILC & \cite{fDfnal} & 215(14) & 254(14) & 1.188(26) \\
\hline
\hline
\end{tabular}
\end{center}
\caption{Comparison of $f_{D^+}$ and $f_{D_s}$ as obtained from experiment 
and from unquenched lattice QCD calculations. \label{tablefD}}
\end{table}
Both lattice collaborations agree very well 
in the central values obtained for the decay constants. 
The errors of the HPQCD calculation are smaller 
than those for the FNAL/MILC due to the fact that the HISQ action 
is more improved\footnote{Improvement in this context refers to 
the addition of higher-dimensional operators to the action.} 
than the Fermilab action and thus discretization errors 
are sensibly reduced. 
The FNAL/MILC results agree with experiment for $f_{D^+}$ and $f_{D_s}$ 
within errors. The HPQCD result for $f_{D^+}$ agrees also very well 
with the experimental number. However there is a discrepancy of 
over $3\sigma$ between the HPQCD and experimental values for $f_{D_s}$. 
This discrepancy has been recently suggested to be a possible 
hint of beyond the Standard Model effects \cite{BK08}. More work is 
needed to resolve this issue. From the experimental side, 
a reduction of the errors that will constrain the possible  
statistical fluctuations is expected. 
It is also desirable that certain issues like the assumption of
three-generation CKM unitarity to set $V_{cs}= V_{ud}$ or the inclusion of
radiative corrections from experimental data or Monte Carlo simulations are
addressed. From the theory side, the error on the FNAL/MILC result for
$f_{D_s}$ is currently larger by roughly a factor of four than the error on
the HPQCD result. The FNAL/MILC collaboration plans to reduce the
uncertainties in their calculation in the near future by increasing
statistics, using a smaller lattice spacing, and improving the determination
of the inputs needed. This will constitute an important check of the HPQCD
numbers.

Calculations of $f_{D^+}$ using Wilson fermions are also making progress, 
although they are still restricted to $N_f=2$ simulations. 
A value of $f_{D^+}=201\pm22^{+4}_{-9}$ 
based on a single value of the lattice spacing was reported in 
\cite{Haas08}. 
The number is compatible with the $N_f=2+1$ calculations 
but with larger errors. The authors in that reference determine the 
ratio $f_{D^+}/f_\pi$ and use the experimental value of $f_\pi$ instead of 
directly determine $f_{D^+}$ since the chiral logarithms are suppressed in 
the ratio and the associated systematic error is thus reduced. 
Preliminary results using twisted mass fermions with $N_f=2$ 
for $f_{D^+}$ and $f_{D_s}$ were also reported by the ETM collaboration 
in \cite{fDETM07} for two different values of the lattice spacing.

\subsection{$B$ and $B_s$ decay constants}

Lattice results for the decay constants in the $B$ sector are more 
needed than in the $D$ sector since the corresponding CKM matrix 
elements to extract the 
information from experiment are worse known. The value of the 
$B$ decay constants are used in the SM predictions for processes very 
sensitive to beyond SM effects, such as $B_s\to\nu^+\nu^-$. The purely 
leptonic decays themselves are also a sensitive probe of effects 
from charged Higgs bosons.

The way of calculating these decay constants on the lattice 
is the same as for the charm decay constants. In fact, the FNAL/MILC 
collaboration has also calculated $f_B$, $f_{B_s}$ and the 
ratio of both with the same choice of actions, same ensembles and 
same procedure as for the $D$ decay constants in \cite{fDfnal}. 
The errors in both the charm and bottom mesons are thus very similar. 
The results are listed in Table 
\ref{tablefB}. In that table, the results from the other lattice 
$n_f=2+1$ calculation, by the HPQCD 
collaboration, are also included. In this case the HPQCD collaboration used 
the NRQCD action mentioned in section \ref{formulations} to 
describe the $b$ valence quark. Errors are then significantly larger 
than in their analysis of $D^+$ and $D_s$ decay constants where they use 
the HISQ action for the $c$ quark. A dominant source of 
uncertainty in their NRQCD calculation is the error associated 
with the one-loop renormalization applied.
\begin{table}[h]
\begin{center}
\begin{tabular}{|c c c c c|}
\hline
\hline
Group & Reference & $f_B$ & $f_{B_s}$ & $f_{B_s}/f_B$ \\
\hline 
FNAL/MILC & \cite{fDfnal} &  197(13) & 240(12) & 1.22(3) \\
HPQCD & \cite{fBhpqcd} & 216(22) & 260(26) & 1.20(3) \\
\hline
\hline
\end{tabular}
\end{center}
\caption{Values of $f_B$ and $f_{D_s}$ as obtained from the  
two lattice QCD calculations with $n_f=2+1$. 
\label{tablefB}
}
\end{table}

A suggested way of reducing the uncertainty in the calculation of 
$f_{B_s}/f_B$ is by extracting it from the double ratio 
$[f_{B_s}/f_B]/[f_K/f_\pi]$ which can be calculated very accurately 
since it is very close to one in ChPT.

\section{Semileptonic decays}

Semileptonic decays can be used to extract CKM 
matrix elements like $\vert V_{cb}\vert$, $\vert V_{ub}\vert$, 
$\vert V_{cd}\vert$, $\vert V_{cs}\vert$ and $\vert V_{us}\vert$. 
The theory input needed to get those 
parameters from experimentally measured semileptonic widths are 
the form factors in terms of which the hadronic matrix elements 
involved on those decays are parametrized. For example, for the decay   
$D\to K l\nu$, the differential decay rate is given by
\be
\frac{d\Gamma}{dq^2} = ({\rm known\, factors})\vert V_{cs}
\vert^2 f_+^2(q^2)\, ,
\ee
where $f_+(q^2)$ is the vector form factor, which can be extracted from the 
matrix element of the vector current
\be
\langle K\vert V^\mu\vert D\rangle  = f_+(q^2)(p_D+p_K-\Delta)^\mu
+f_0(q^2)\Delta^\mu\, ,
\ee
with $\Delta^\mu=(m_D^2-m_K^2)q^\mu/q^2$.

Lattice QCD can be used to calculate the value of those form factors 
as a function of the virtual W momentum transfer, $q^2$, or, equivalently, 
the recoil momentum of the daughter meson. On the lattice, the smallest 
discretization errors correspond to the form factor at the largest 
momentum transfer, where the experimental data are less precise. 
In addition, the finite volume provides an infrared 
cutoff and there is a finite minimum value for the momenta that
can be simulated. A way of 
circumventing this limitation is by using twisted boundary conditions 
that allow for arbitrary small values of the momenta \cite{twboundary}. 
Another set of important techniques that can be applied to semileptonic 
decay analysis are double ratio methods \cite{doubleratios}.  
These methods can yield a reduction of both statistical and systematic 
uncertainties by a partial or total cancellation of those uncertainties 
between numerator and denominator in the ratios.

In the case of $\vert V_{us}\vert$, experimental data 
for $K\to \pi l\nu$ and lattice results for the corresponding 
form factors at zero momentum transfer yield a determination 
of this parameter with an error claimed to be 0.5~\% -see F.~Mescia's 
\cite{mesciaFPCP08} and P.~Gambino's \cite{gambinoFPCP08}
talks at this conference. 

\begin{figure}[h]
\includegraphics[width=0.35\textwidth,angle=-90]{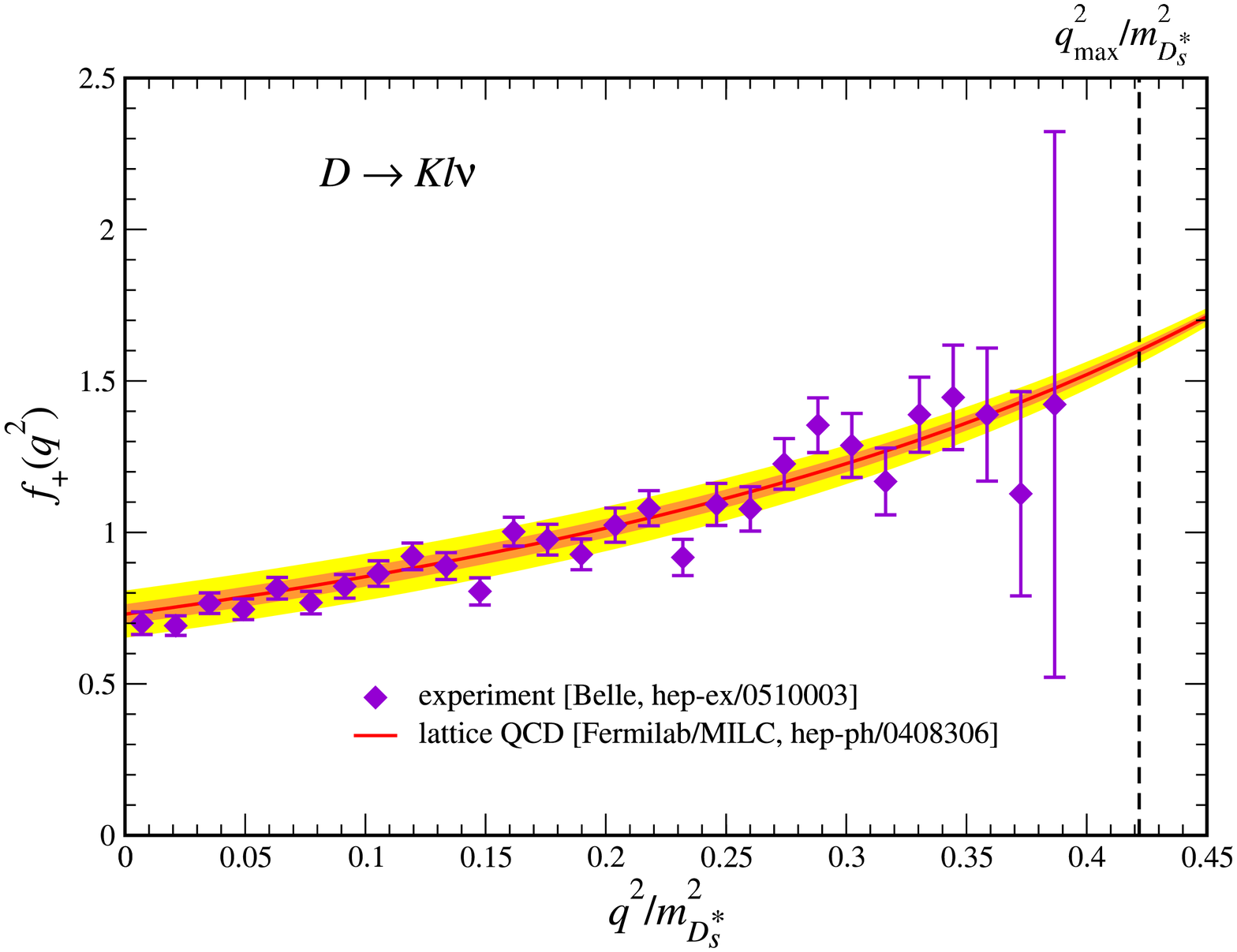}
\caption{Comparison of the shape of the vector form factor $f_+^K(q^2)$ 
as measured experimentally and obtained on the lattice \cite{Kronfeld05}.
\label{DKshape}}
\end{figure}

The FNAL/MILC calculation of the shape of the form factor $f_+^{K}$ 
for the decay $D\to K l \nu$ \cite{DtoKFNAL} in 2004 constituted a prediction 
since the result appeared before the experimental measurements by 
the FOCUS \cite{Focus} and Belle \cite{Belle} collaborations. The 
comparison of this lattice calculation with Belle experimental data is 
shown in Figure \ref{DKshape} as a function of $q^2$. 
The determination of the vector form factor for $D\to K l \nu$ and 
$D\to \pi l \nu$ are being currently updated by the FNAL/MILC 
collaboration. The most straightforward improvement is the extension of 
the calculation to smaller values of the lattice spacing. This will 
reduce discretization errors which are the main source of 
uncertainty in the previous calculation.

The matrix element $\vert V_{cb}\vert$ can be extracted from the 
decay $B\to D^* l \nu$. The experimental results for this process at 
zero recoil have smaller errors than those for $B\to D l \nu$. 
A new lattice calculation of the form factor  
describing the decay, the axial vector form factor at zero recoil 
${\cal F_{B\to D^*}}(1)$, was presented in the lattice conference last year 
\cite{LaihoLat07}. This calculation eliminates the quenching errors from 
previous calculations since it includes $n_f=2+1$ sea quarks. 
It also introduces a new double ratio method which gives the form 
factor at zero recoil directly and with a reduction of the 
computational cost. The relation between the double ratio and the 
form factor, 
\be
\vert {\cal F_{B\to D^*}}(1) \vert^2 = \frac{\langle D^*\vert \bar c 
\gamma_j\gamma_5 b\vert\bar B\rangle \langle \bar B\vert \bar b 
\gamma_j\gamma_5 c\vert D^*\rangle }
{\langle D^*\vert \bar c \gamma_4 c\vert D^*\rangle 
\langle \bar B\vert \bar b \gamma_4 b\vert\bar B\rangle }
\, ,
\ee
is  exact to all orders in the heavy-quark 
expansion in the continuum. Statistical errors in the numerator and 
denominator are highly correlated and largely cancel. And also, most 
of the renormalization cancels, yielding a small uncertainty for the 
perturbative matching.

The final result obtained for the form factor after chiral and 
continuum extrapolation is ${\cal F}_{B\to D^*}(1)=0.921
\pm0.013\pm0.021$ \cite{Laiho08}, where the first error is 
statistical and the second one includes all sources of systematic 
errors and is dominated by heavy-quark discretization errors. The 
CKM matrix element $\vert V_{cb}\vert$ extracted from 
this value of the form factor and the experimental averages in 
\cite{HFAG08} is
\be
\vert V_{cb}\vert = (39.2 \pm 0.6 \pm 1.0)\times 10^{-3}\, .
\ee
This value differs by $2\sigma$ from the one extracted from 
inclusive decays.

An alternative method for the extraction of $\vert V_{cb}\vert$ is the 
analysis of the decay $B\to D l\nu$. A study of the form 
factors needed for such determination with quenched Wilson 
fermions was presented in \cite{Divitiis07}. An interesting
aspect of that analysis is that the authors calculate the scalar
form factor as well as the momentum transfer dependence  
to avoid needing to extrapolate to zero recoil, 
where the experimental data suffer from phase space
suppression compared to the $B\to D^* l\nu$ case. The scalar form factor, 
which only contributes for $l=\tau$ can be used to constrain BSM 
physics \cite{KM08}. 

\begin{figure}[h]
\includegraphics[width=0.35\textwidth,angle=-90]{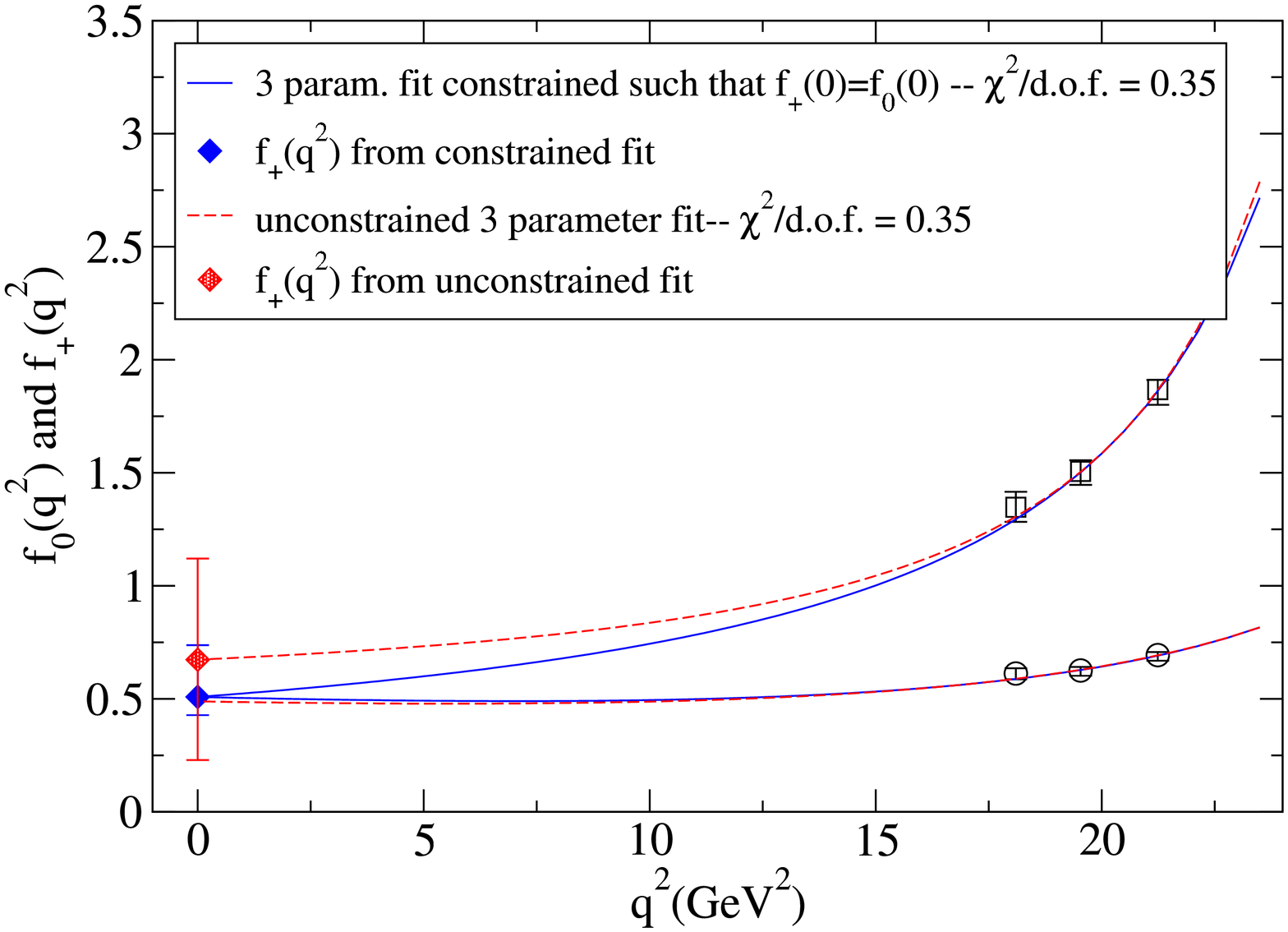}
\caption{Preliminary results in \cite{WMLat07} for the form factor 
describing the decay $B\to\pi l\nu$ as a function of $q^2$.
\label{figBpi}}
\end{figure}

The decay $B\to \pi l\nu$ provides a way of extracting $\vert V_{ub}\vert$ 
that is competitive with $b\to u$ inclusive decays. The only 
unquenched ($N_f=2+1$) lattice determination of the form factor 
needed for this extraction to date is the HPQCD calculation in 
\cite{HPQCDformfactors}, which uses the Asqtad staggered formulation for 
the light quarks and NRQCD for the $b$ quarks. Together with experimental 
results, from this calculation one obtains \footnote{First error is 
experimental, the second one is the theoretical error from 
the lattice calculation of the vector form factor.} 
$\vert V_{ub}\vert=(3.55\pm0.25\pm0.50)\times 10^{-3}$. 
The error is larger than it could be due to the fact that lattice 
and experimental results for the corresponding decay rate  have 
a poor overlap in $q^2$. 
There are several methods that can be adopted to try to address this issue.
The HPQCD collaboration is reducing the recoil
momentum of the pion required for small $q^2$ by using a lattice frame in
which the $B$ meson is moving in the opposite direction to the pion. A
modified version of NRQCD (moving-NRQCD) provides the description of the
corresponding b quark with a large velocity \cite{WongLat07}. It is then
possible to calculate the form factors at small $q^2$ without needing large
recoil momenta, hence keeping discretization and statistical errors under
control. The FNAL/MILC collaboration is implementing another method to
overcome the poor overlap \cite{WMLat07}. They are using the so
called $z-$expansion to parameterize the shape of  the form factors. This
parameterization is  model independent and
based only on unitarity and analyticity \cite{zexp}.  When the experimental
data are analyzed using the same model-independent parameterization, the
results for the shape parameters from theory and experiment can be compared
directly even if the respective $q^2$ ranges have poor overlap. Hence,
Lattice QCD  must provide only the normalization of the form factor, which
can be calculated at the $q^2$ values where theoretical errors are smallest.
The FNAL/MILC collaboration uses Fermilab action for the $b$ quark and
improved staggered formalism to describe the light quark in the numerical
simulations for this work. The preliminary results obtained by this 
collaboration are shown in Figure \ref{figBpi}. They look quite promising. 

The calculation of semileptonic and leptonic decays on the lattice 
can be used to construct ratios independent of CKM matrix elements, 
such as $\frac{\Gamma(D\to l\nu)}{\Gamma(D\to \pi l\nu)}$ or  
$\frac{\Gamma(D_s\to l\nu)}{\Gamma(D\to K l\nu)}$, 
with which one could test the consistency of lattice calculations 
against experiment or constrain BSM physics.

\section{Neutral meson mixing}

The theoretical determination of the parameters that describe the 
mixing in the neutral Kaon and $B$ systems are needed in order to test 
for the consistency of the SM description of CP violation when 
compared against experiment. 

In the SM, the neutral $K$ and $B$ meson mixing is due to box diagrams with 
exchange of two $W$-bosons. These box diagrams can be rewritten in terms of 
an effective Hamiltonian with four-fermion operators describing 
processes with $\Delta F=2$ (F=S or B). The matrix elements 
of the operators 
between the neutral meson and antimeson encode the non-perturbative 
information on the mixing and can be calculated using lattice QCD 
techniques.

\subsection{Indirect CP violation in neutral kaon decays: $B_K$.}

The non-perturbative input to study CP violating effects in 
$K^0-\bar K^0$ mixing is parametrized by $B_K$, defined as
\be
B_K(\mu)\equiv \frac{\langle\bar K^0|
{Q_{\Delta S=2}(\mu)}|K^0\rangle}
{\frac{8}{3}\langle \bar K^0|\bar s\gamma_{\mu}\gamma_5d|0\rangle
\langle 0|\bar s\gamma_{\mu}\gamma_5d|K^0\rangle}\,.
\ee
The theoretical calculation of this parameter, together with the 
experimental measurement of $\varepsilon_K\equiv 
\left\vert\frac{A(K_L\to \left(\pi\pi\right)_{I=0})}
{A(K_S\to \left(\pi\pi\right)_{I=0})}\right\vert$, gives a hyperbole in 
the $\rho-\eta$ plane, where $\rho$ and $\eta$ are the usual unitarity 
triangle parameters. That corresponds to the light blue band 
in Figure \ref{ckmplot}. 

\begin{figure}[h]
\includegraphics[width=0.4\textwidth,angle=-90]{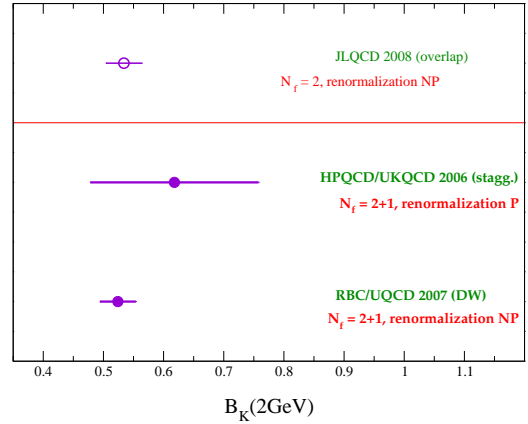}
\caption{Recent unquenched lattice values of $B_K$. The parameter 
is given in the $\overline{MS}-NDR$ scheme and at an scale equal 
to $2{\rm GeV}$. The results are taken from 
\cite{overlapBK} (JLQCD), \cite{ourBK} (HPQCD/UKQCD) and 
\cite{BKdw} (RBC/UKQCD). The matching is done nonperturbatively 
in \cite{overlapBK} and \cite{BKdw}, and perturbatively at one-loop 
in \cite{ourBK}. The perturbative matching is the origin of the 
rather large error. \label{BKresults}}
\end{figure}

The most recent unquenched lattice determinations of $B_K(\mu)$ are shown 
in Figure \ref{BKresults}. 
The current most accurate value is the one by the RBC/UKQCD collaboration 
using domain wall fermions with $n_f=2+1$ sea quarks \cite{BKdw}
\be\label{BKdwvalue}
B_K^{\overline{MS}}(2~{\rm GeV})=0.524\pm0.010\pm0.013\pm0.025\,,
\ee
where the first error is statistical, the second one is associated 
with the non-perturbative renormalization and the third one corresponds 
to the other systematic errors. The main source of uncertainty in this 
calculation is discretization errors. The result in (\ref{BKdwvalue}) 
is obtained from simulations with a single value of the lattice spacing,  
$a^{-1}=1.729(28){\rm GeV}$. The discretization error 
corresponding to the use of this single value of $a$ 
is estimated in \cite{BKdw} by using the scaling behaviour of a previous 
quenched calculation with the same light quark action. The authors found 
that error to be of around a 4\%, which dominates 
the systematic uncertainty. The RBC/UKQCD collaboration is planning to 
explicitly check discretization effects by performing 
the unquenched calculation with the same quark and gluon actions  
but with a smaller lattice spacing. 

The JLQCD result in Figure \ref{BKresults} with $N_f=2$ sea quarks is 
affected by the fact that part of the vacuum polarization effects are 
still missing and are difficult to estimate \emph{a priori}. 
This collaboration 
is planning on extending its methodology to a $N_f=2+1$ calculation of 
$B_K$. The needed configurations are being currently generated. The 
ensemble will include also configurations for two different volumes.  
This will allow them to explicitly study and reduce the finite volume 
effects that are the main source of uncertainty in their $N_f=2$ 
calculation.

There are two other unquenched determinations of $B_K$ in progress which 
use mixed actions\footnote{In a mixed action calculation the valence 
and sea quarks are described with different fermion formulations.}. A  
$N_f=2$ calculation outlined in \cite{ScorzatoLat07} uses overlap 
fermions for the valence quarks and twisted mass fermions for the sea quarks. 
A $N_f=2+1$ study whose preliminary results can be found in 
\cite{BKRuthJack} uses domain wall valence fermions and improved 
staggered sea fermions. The errors due to the matching to the continuum 
affecting previous staggered calculations can be highly 
reduced in this analysis by performing the renormalization 
non-perturbatively. The use of domain wall valence quarks makes the 
chiral extrapolation more continuum-like than in the purely staggered 
case. Another advantage of this calculation is that there are  
staggered configurations generated by the MILC collaboration for a 
large range of lattice spacings, volumes and small sea quark masses. 
This allows for good control over the systematic error from chiral 
and continuum extrapolations. The expected final error from 
this calculation is around $5\%$. 

In the near future, there will be thus several lattice calculations 
of $B_K$ using different discretizations and with errors at the $5\%$ level.

\subsection{$B^0$ mixing: $\Delta M_{d,s}$, $\Delta \Gamma_{d,s}$ and 
$\xi$}

\label{B0mixing}

The mixing in the $B^0_q-\bar B^0_q$ system is an interesting  
place to look for NP effects. The BSM effects can appear 
as new tree level contributions, or through the presence of new particles 
in the box diagrams. In fact, it has been recently claimed that there is 
a disagreement 
between direct experimental measurement of the phase of $B^0_s$ mixing 
amplitude and the SM prediction \cite{UTfit08}. Possible NP effects have also 
been reported to show up in the comparison between direct experimental 
measurements of $\sin(2\beta)$ and SM predictions using $B^0$ 
mixing parameters \cite{LunguiSoni08}. Studies of neutral $B$ meson 
mixing parameters can also impose important constraints on different 
NP scenarios \cite{fleischer}. 

The quantities that describe the mixing in the $B^0$ system are the 
mass differences,  $\Delta M_{s,d}$, and decay width differences, 
$\Delta \Gamma_{s,d}$, between the heavy and light $B^0_s$ and 
$B^0_d$ mass eigenstates. The non-perturbative physics of those processes 
is contained in hadronic matrix elements of the four-fermion operators 
in the effective Hamiltonian with $\Delta B=2$. Those matrix elements  
are parametrized by products of $B$ decay constants and bag parameters. 
For example, for the mass difference
\be\label{SMMsd}
\Delta M_{s(d)}\vert_{theor.}\propto
\vert V_{t s(d)}^*V_{tb}\vert^2
f_{B_{s(d)}}^2\hat B_{B_{s(d)}}\, ,
\ee
with  $\langle \bar B^0_s\vert Q^{s(d)}_L \vert B^0_s
\rangle = \frac{8}{3}M^2_{B_{s(d)}}f^2_{B_{s(d)}}
B_{B_{s(d)}}(\mu)$ and $O^{s(d)}_L = \left[\bar b^i 
\gamma_\mu(1-\gamma_5)s^i(d^i)\right]\,
\left[\bar b^j\gamma^\mu(1-\gamma_5)s^j(d^j)\right]$. 

Many of the uncertainties that affect the theoretical calculation
of the decay constants and bag parameters cancel totally or partially 
if one takes the ratio $\xi^2=f_{B_s}^2 B_{B_s}/f_{B_d}^2 B_{B_d}$. 
Hence, this ratio and therefore the combination of CKM matrix elements 
related to it, $\vert \frac{V_{td}}{V_{ts}}\vert$,  
can be determined with a significantly 
smaller error than the individual matrix elements. The ratio $\xi$ is also 
an important ingredient in the unitarity triangle analyses.

The first lattice calculation of the $B^0$ mixing parameters with 
$n_f=2+1$ sea quarks, which only studied the $B^0_s$ sector, was 
performed by the HPQCD collaboration in \cite{ourbsmix}. The authors 
obtained 
\be\label{masaresult}
\Delta M_s = 20.3(3.0)(0.8)ps^{-1}\quad{\rm and}\quad
\Delta \Gamma_s = 0.10(3)ps^{-1}\,, 
\ee
which is compatible with experiment. 

The FNAL/MILC and HPQCD collaborations are currently working on a more 
complete study of $B^0-\bar B^0$ mixing, including $B^0_s$ and 
$B^0_d$ parameters. The main goal of both projects is obtaining the ratio 
$\xi$ fully incorporating vacuum polarization effects. 
The choice of actions and the setup is the same as for their $f_B$, 
$f_{B_s}$ calculations described in Section \ref{leptonic} 
-more details can be found in \cite{B0mixingproceedings}. 

\begin{figure}[h]
\includegraphics[width=0.35\textwidth,angle=-90]
{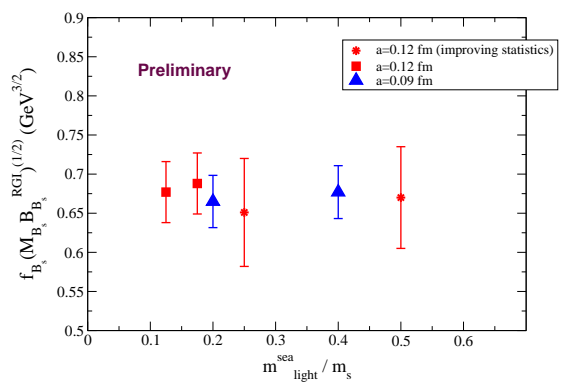}
\caption{Values of $f_{B_s}\sqrt{M_{B_s}\hat B_{B_s}}$ in ${\rm GeV}^{3/2}$ 
as a function 
of the light sea quark mass normalized to the physical strange quark mass 
from the HPQCD collaboration. The data include statistical, 
perturbative and scale errors. The bottom valence quark is fixed 
to its physical value and the strange valence quark is very close 
to its physical value. The strange sea quark mass is also very close to its 
physical value.\label{fBBBHPQCD}}
\end{figure}

\begin{figure}[h]
\includegraphics[width=0.35\textwidth,angle=-90]{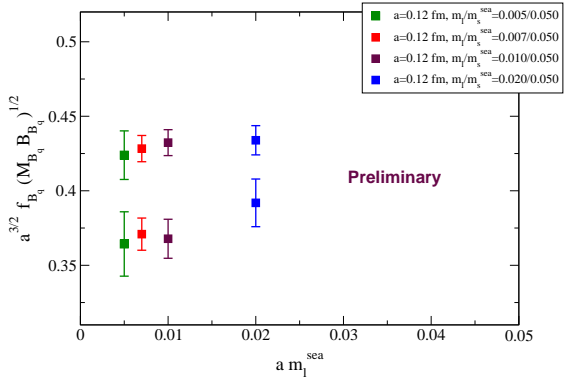}
\caption{Bare values of $f_{B_q}\sqrt{M_{B_q} B_{B_q}}$ in lattice units 
as a function of the light sea quark mass also in lattice units from 
the FNAL/MILC collaboration. Results are shown for both $B^0_s$ 
and $B^0_d$ including only statistical errors. The results correspond  
to one of the three lattice spacings at which the  FNAL/MILC's study 
is performed. The bottom valence quark is fixed 
to its physical value and the strange valence quark is very close 
to its physical value. The strange sea quark mass is also very close to its 
physical value.
\label{fBBBFNAL}}
\end{figure}

Results from the two collaborations are still preliminary. Figures  
\ref{fBBBHPQCD} and \ref{fBBBFNAL} show some examples of 
the values of $f_B\sqrt{M_B B_B}$ obtained as function of the light sea 
quark masses. 
The dependency on the light sea quark mass is in both studies very mild, 
so only the chiral extrapolation in the $d$ quark mass for $B^0_d$ 
parameters is expected to be a significant source of error. 
In Figure \ref{fBBBHPQCD}, it can also 
be appreciated that the results for the two different lattice spacings 
are very similar, which indicates small discretization errors. 

\begin{figure}[h]
\begin{center}
\includegraphics[width=0.35\textwidth,angle=-90]{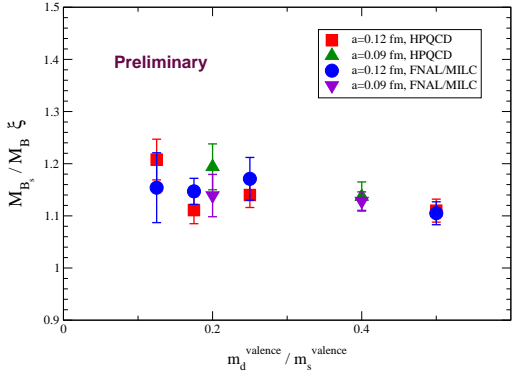}
\end{center}
\caption{Product of the ratios $\xi$ and $M_{B_s}/M_{B_d}$ as 
a function of the down quark mass normalized to the strange quark mass. 
Results for both FNAL/MILC and HPQCD collaborations including only 
statistical errors are shown for two different values of the lattice 
spacing $a$. 
\label{xicomparison}} 
\end{figure}

A comparison of the preliminary results from the two collaborations 
for the ratio $\xi$ is shown in Figure \ref{xicomparison}.
Only the full QCD points\footnote{Where valence and sea quark 
masses are the same.} are included. The results of the two 
collaborations agree within statistical 
errors. This is very encouraging since both analyses use completely 
different descriptions for the heavy quarks.

The final step in those calculations is extrapolating the results 
to the physical values of the light quark masses and, in the case of 
the FNAL/MILC collaboration, performing simultaneously the extrapolation 
to the continuum. Systematic errors on those extrapolations are currently 
being studied. These two analyses are expected to have final results 
very soon with total errors ranging $5-7\%$ for $f_{B_q}\sqrt{B_{B_q}}$ and 
$2-3\%$ for $\xi$. 

The effects of heavy new particles in the box diagrams that describe 
the $B^0$ mixing can be seen in the form of effective 
operators built with SM degrees of freedom. The NP could modify the 
Wilson coefficients of the four-fermion operators that already contribute 
to $B^0$ mixing in the SM and gives rise to new four-fermion operators in 
the $\Delta B=2$ effective Hamiltonian -see \cite{gabbiani,damir} for a list 
of the possible operators in the SUSY basis. The calculation of 
those Wilson coefficients for a particular BSM theory, together with 
the lattice calculation of the matrix elements of all the possible 
four-fermion operators in the SM and beyond and experimental 
measurements of $B^0$ mixing parameters, can constraint BSM 
parameters and help to understand new physics. To date, there does not 
exist an unquenched determination of the complete set of matrix elements 
of four-fermion operators in that general $\Delta B=2$ effective Hamiltonian. 
However, the two collaborations currently working on $B^0$ mixing in 
the SM are planning to extend their analysis to BSM operators in the near 
future. Actually, the HPQCD collaboration has already calculated 
the one-loop matching coefficients needed for such an analysis 
\cite{Bb_bsm}.

\section{Conclusions}

Hints of discrepancies between SM predictions and experimental 
measurements have started to show up in some CP violating 
observables \cite{BG08}. As claimed in \cite{BG08}, 
the precise determination of parameters like $\hat B_K$, $f_K$ and $\xi$, 
and CKM matrix elements like $\vert V_{cb}\vert$  
is crucial in order to fully exploit 
the potential of CP violating observables on constraining NP. 
Lattice QCD has a fundamental role in that program. 
There has recently been important progress  
in order to achieve  realistic lattice calculations, with $n_f=2+1$ 
sea quarks and a serious study of systematic errors. New results relevant 
for phenomenology have appeared in the last year 
in the Kaon and $D$ meson sectors with errors at the few percent level. 
In the near future, results for $B^0$ mixing parameters and 
$B$ decay constants will be also available with errors at the few percent 
level.

\begin{figure}
\includegraphics[width=7.0cm,height=7.0cm]{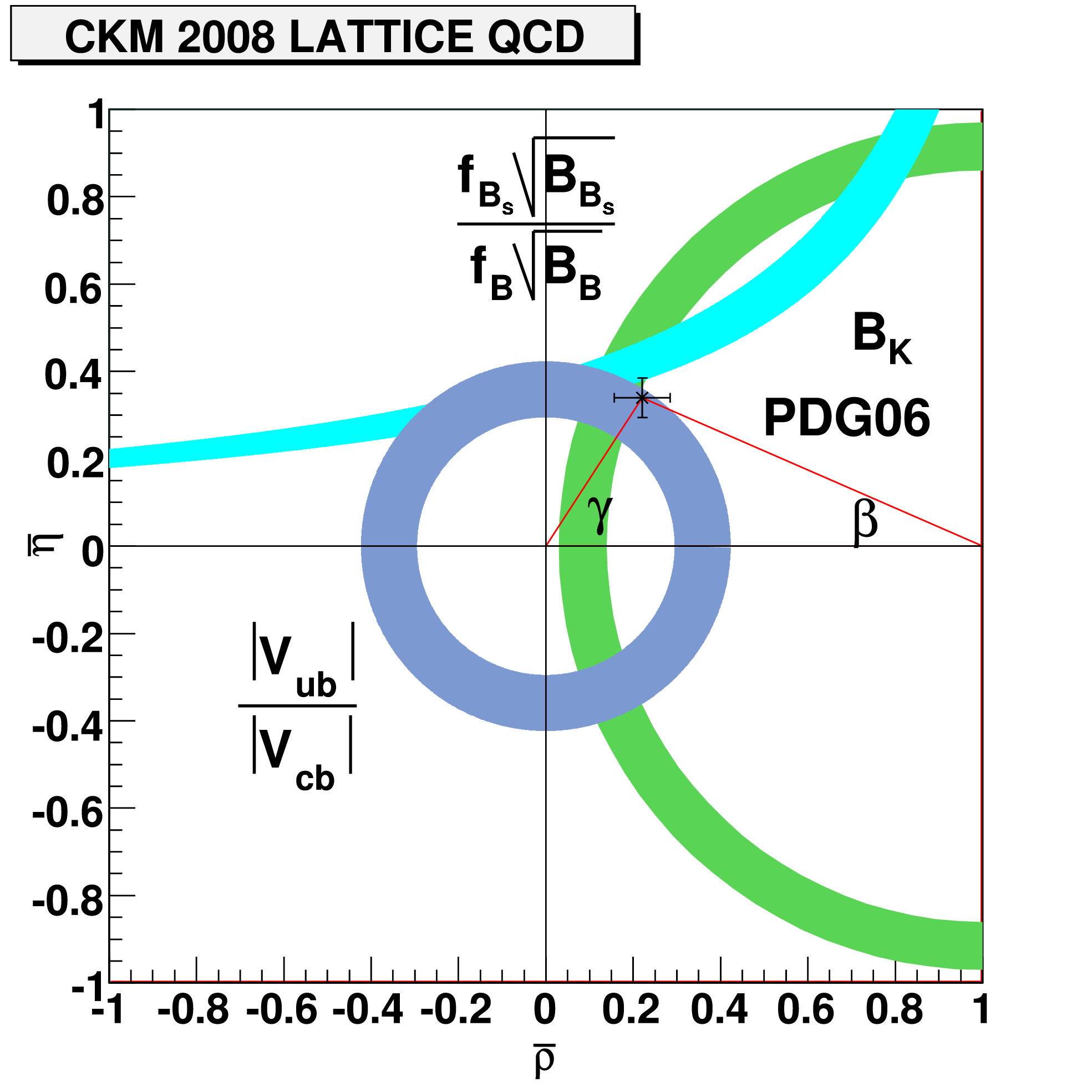}
\caption{Constraints on the $\bar\rho-\bar\eta$ plane 
imposed by recent unquenched lattice QCD calculations. 
The black point and error bars correspond 
to the values of $\bar \rho$ and $\bar \eta$ from the PDG 
2006 review of particle physics \cite{PDG06}. 
See the text for more explanations. The plot is an updated version of 
the one in \cite{Christinelp}. \label{ckmplot}}
\end{figure}

Figure \ref{ckmplot} summarizes the impact of the lattice calculations 
with $n_f=2+1$ on the unitarity triangle 
analysis. In generate the plot, the value of $B_K$ is the one by 
the RBC/UKQCD collaboration \cite{BKdw}, 
$V_{us}$ is taken from leptonic decays with $f_K/f_\pi$ given by the 
HPQCD collaboration \cite{fDhpqcd}, $\vert V_{cb}\vert$ is from 
semileptonic $B\to D^* l \nu$ with the form factor by the FNAL/MILC 
collaboration \cite{Laiho08},  $\vert V_{ub}\vert$ is from 
Flynn and Nieves \cite{FN07} using, among other information, 
the form factor for $B\to\pi l\nu$ by the HPQCD \cite{HPQCDformfactors}  
and FNAL/MILC \cite{FpFNAL,WMLat07} collaborations. 
A final $n_f=2+1$ result for $\xi$ is not yet available. In order to
illustrate its effect on the $\rho-\eta$ plane in Figure 8, 
we assumed a value for $\xi$ with a
3\% error. As explained in Section \ref{B0mixing}, this is the expected
error for $\xi$ from both the HPQCD and the FNAL/MILC calculations.

Several lattice collaborations are currently producing accurate
$n_f=2+1$ results as discussed in this paper, 
and other collaborations are starting to generate 
$n_f=2+1$ ensembles using different sea quark actions. This will allow 
an important consistency check of lattice methods.

\begin{acknowledgments}

I thank Aida El-Khadra, Christine Davies, Junko Shigemitsu, Eduardo Follana, 
Ruth Van de Water, Jack Laiho, Todd Evans and Claude Bernard  
for help in the preparation of the talk. I thank the organizers 
to invite me to this enjoyable meeting. 
This work was supported in part by the DOE 
under grant no. DE-FG02-91ER40677, by the Junta de Andaluc\'{\i}a 
[P05-FQM-437 and P06-TIC-02302] and by URA Visiting Scholars Program.  

\end{acknowledgments}

\bigskip % extra skip inserted

\end{document}